# Design Optimization of Flip FET Standard Cells with Dual-sided Pins for Ultimate Scaling

Rui Guo, Haoran Lu, *Student Member, IEEE*, Jiacheng Sun, *Student Member, IEEE*, Xun Jiang, Lining Zhang, *Senior Member, IEEE*, Ming Li, *Member, IEEE,* YiBo Lin, *Member, IEEE,* Runsheng Wang, *Member, IEEE*, Heng Wu, *Member, IEEE* and Ru Huang, *Fellow, IEEE*

*Abstract*— Recently, we proposed a novel transistor architecture for 3D stacked FETs called Flip FET (FFET), featuring N/P transistors back-to-back stacked and dual-sided interconnects. With dual-sided power rails and signal tracks, FFET can achieve an aggressive 2.5T cell height. As a tradeoff, the complex structure and limited numbers of M0 tracks could limit the standard cell design. As a solution, multiple innovations were introduced and examined in this work. Based on an advanced node design rule, several unique building blocks in FFET such as drain merge (DM), gate merge (GM), field drain merge (FDM) and buried signal track (BST) were investigated. Other key design concepts of multi-row, split gate and dummy gate insertion (DG) were also carefully studied, delivering around 35.6% area reduction compared with 3T CFET. Furthermore, the symmetric design of FFET has unique superiority over CFET thanks to the separate N/P logic on two sides of the wafer and their connections using DM and GM. New routing scheme with dual-sided output pins on both wafer frontside (FS) and backside (BS) was proposed for the first time. Finally, we conducted a comprehensive evaluation on complex cell design, taking AOI22 as an example. New strategies were proposed and examined. The FDM design is identified as the best, outperforming the BST and dummy gate design by 1.93% and 5.13% for the transition delay.

## I. Introduction

With the continuous scaling of electronic devices [1], a new architecture, the 3D stacked transistor [2][3] was proposed to continue Moore's Law. Under this trend, CFET [4]-[8] is considered to be one of the candidates for the next generation advanced logic device. With nFET stacked vertically over pFET, CFET achieves great standard cell (std. cell) area reduction by shrinking the cell height down to 3T [9]. However, CFET faces many challenges, such as high aspect ratio processes, high intrinsic parasitic RC and high pin density. All input and output pins of CFET are placed on the frontside (asymmetric std. cell design), which brings great challenges in the std. cell design. Moreover, the placement and routing (P&R) of CFET is large because of the high pin density on the frontside [10], even if the local backside interconnect is introduced. Considering this, more area may be wasted to enable the chip design with ultra-scaled 3T CFET cell design due to the routing congestion and the limitation of the common gate design. Naturally, for the stacked FET std. cell design, the symmetric design is preferred, i.e., N logic (power & signal) placed on the one side while P logic (power & signal) on the other side. As a result, we recently proposed Flip FET (FFET) [11], a novel 3D stacked transistor architecture with vertically stacked transistors in the back-to-back-stacking fashion. This architecture enables dual side BEOL and equally distributes the pins to both sides, which brings more signal tracks and better routability. Thus, FFET can achieve a minimum 2.5T cell height. Fig. 1(a) shows the basic layout of FFET inverter with one driven strength (INVD1). The frontside and backside layers are separately shown in this and the following figures. 2 signal and 1 power tracks are placed on each side, which brings one more signal track than 3T CFET. Different cross-sections of FFET with critical dimensions highlighted are also given in Fig.

This work was supported in part by the National Key R&D Program of China under Grant 2023YFB4402200; in part by the 1+1 Project under Grant QYJS-2023-2301-B; in part by the NSFC under Grant 92464206; and in part by Beijing Outstanding Young Scientist Program (JWZQ20240101004)

The authors are with the School of Integrated Circuits, Peking University, Beijing 100871, China. Rui Guo is also with the School of Software & Microelectronics, Peking University, Beijing 102600, China. (e-mail: hengwu@pku.edu.cn).

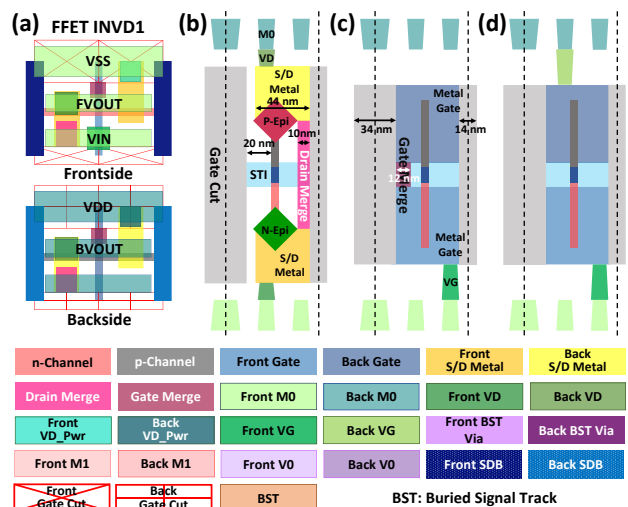

Fig. 1. (a) The layout of FFET INVD1. (b)-(d) Cross-section of FFET's drain region with Drain Merge (DM), gate region with Gate Merge (GM) and without Gate Merge. Some critical dimensions are also given.

1. Detailed dimensions are shown in Table 1. Note that we choose single fin instead of stacked nanosheets for FFET's channel structure in this work, focusing on aggressively scaled standard cell designs.

In this work, we designed the standard cell libraries of 2.5T FFET and 3T CEFT consisted of 28 different cells. We proposed several design methods and introduced several new building blocks, such as Field Drain Merge (FDM) and Buried Signal Track (BST). Besides, we compared the FFET standard cells area with CFET (Fig. 2). The results indicate that the FFET could have more scaling potential than the CFET.

## II. Key Building Blocks of Standard Cell Design

Considering the dual-sided structure of FFET, certain electrodes of frontside and backside transistors need to be connected to transmit signals accordingly, requiring some unique structures. The first one is the Drain Merge (DM), which connects the frontside and backside S/D metals with a tall via. DM can be formed in a similar fashion to the dual-damascene process. Backside S/D metal is first patterned on the hard mask. Then DM via is patterned and etched, stopping on the frontside S/D metal. After that, backside S/D metal trench is etched to epi, followed by the liner deposition, metal fill and CMP. Lastly, S/D metal fills in both the S/D metal trench and the DM via.

The second one is the Gate Merge (GM), which connects the frontside and backside gates. After the backside RMG and backside metal gate cut, a deep via etch through the metal gate and the STI is conducted, followed by the via metallization.

DM and GM are the most basic building blocks in FFET standard cell design and are used very frequently. The use of DM and GM needs to be carefully treated considering the tradeoff between parasitic resistance and capacitance. Furthermore, they also require further process development to form the high aspect ratio vias.

Besides, these two options have a common issue: they can only be used to connect two S/D metals or gates which are directly stacked on top of each other. For those S/D metals or gates which are vertically staggered, the third option called Field Drain Merge (FDM) is needed. Fig. 3(a) is a sample layout which consists of both FDM and none FDM regions. The FDM is introduced in the field region with the fin removed (Fin Cut), to connect the frontside and backside M0s with a DM, as shown in Fig. 3(b), while the cross-section of the none FDM region is shown in Fig. 3(c). Fig. 4 shows the process flow of FDM and the region without FDM is also given for better comparison. After the self-aligned fin patterning, the frontside fin in FDM region is removed by the Fin Cut process, followed by the frontside devices formation. Then the wafer is flipped and grinded, the backside fin in FDM region is also removed. The backside devices are then formed and whole process is done. What's more, the Fin Cut process can also be done from the frontside and remove the entire active fin at one time. Now the vertically staggered S/D metals can firstly be picked up by M0, then connected together by FDM. However, an extra CPP is wasted due to the extra field region created for connection.

The fourth option is called the Buried Signal Track (BST), a unique structure in FFET. In the FFET, the frontside and backside devices share a common STI region, in which a buried metal can be inserted and used as an extra local M0 line to transmit signals between the S/D metals and gates. The process flow of the BST is similar to the buried power rail (BPR) [12]. After STI formation, a metal track with a thickness of 20 nm is buried in the middle of the deep STI with a TiN liner wrapped around, which can help reduce the diffusion of BST metal. Then frontside fins (or nanosheet stacks) are revealed and the frontside devices are fabricated. Afterwards, the wafer is flipped, and finally the backside devices are formed. Note that the BST may face metal contamination concerns, similar to that of BPR. There are several potential solutions: use TiN liner, use double flips process [13], use W or Ru instead of Cu. The BST can pick up signals from either frontside or backside, as shown in Fig. 5. Even better, it consumes no additional area thanks to that it's buried in the STI within the cell. Fig. 5(c) shows the area comparison of FFET with and without BST. The results indicate that using BST can further reduce the area by 4.4%.

TABLE I
Detailed dimensions of FFET

| Type | Name | Size/nm | Description |
|---|---|---|---|
| Fin | Fin CD | 6 | fin width |
| MD | MD CD | 20 | S/D metal width |
| Gate | Gate Pitch | 51 | gate pitch |
| | Gate Extension (P) | 20 | gate extension at power side |
| | Gate Extension (S) | 25 | gate extension at signal side |
| GM | GM CD | 12 | size of gate merge |
| | GM to Fin Space | 8 | gate merge to fin space |
| DM | DM CD | 16 | width of drain merge |
| | DM Height | 20 | height of drain merge |
| | DM to Fin Space | 5 | drain merge to fin space |
| Via | VD CD | 14 | size of Via Drain |
| | VG CD | 12 | size of Via Gate |
| | VD_power CD | 16 | size of Via Drain power |
| | V0 CD | 14 | size of Via 0 |
| BST | BST CD | 22 | width of BST metal |
| | BST Thk. | 20 | thickness of BST metal |
| BEOL Metal | M0 CD | 14 | width of M0 line |
| | M0 Space | 14 | space of M0 line |
| | M1 CD | 20 | width of M1 line |
| | M1 Space | 14 | space of M1 line |

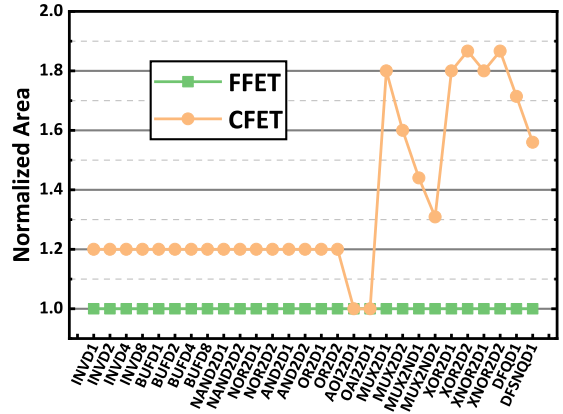

Fig. 2. Normalized area comparison between FFET and CFET standard cells.

Fig. 3. (a) Sample layout of FDM. (b) Cross-section of FDM region. (c) Cross-section of none FDM region.

What's more, the FFET has a unique feature compared with other stacked transistor architectures. The gates in FFET are naturally separated for the wafer frontside and backside, making the Split Gate (SG) structure very easy to realize, while the CFET needs much more effort to realize the SG [10][14]. The simple SG design in the FFET can greatly save area, as shown in Fig. 2. More details will be discussed later.

Last but not least, inserting dummy gates and the multi-row design can also be applied in FFET's layout design. A dummy gate can split the source or drain of two adjacent transistors, and the multi-row design can be used in complex cells design where lots of interconnects are needed.

## III. LAYOUT DESIGN

Due to the symmetric design methodology and the dual-sided routing rule, the layouts of FFET are completely different from CFET and other device structures. In this chapter, we will discuss various methods of FFET's layout design in detail.

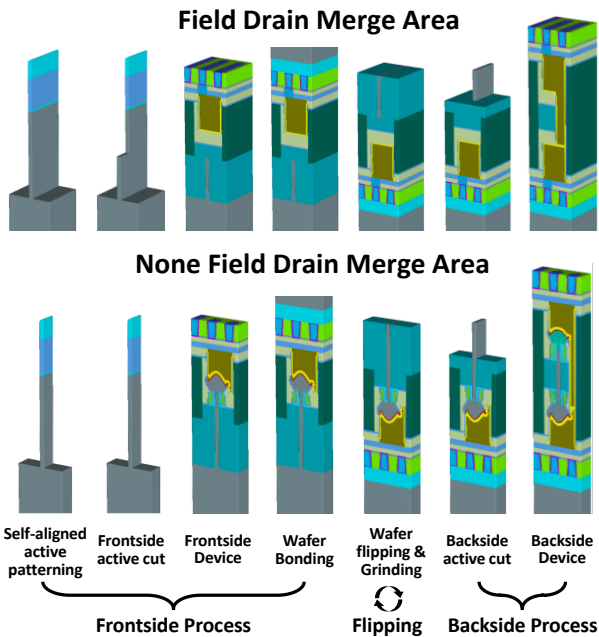

Fig. 4. Process Flow of FDM. Using fincut to create a field region, which can be used for drain merge.

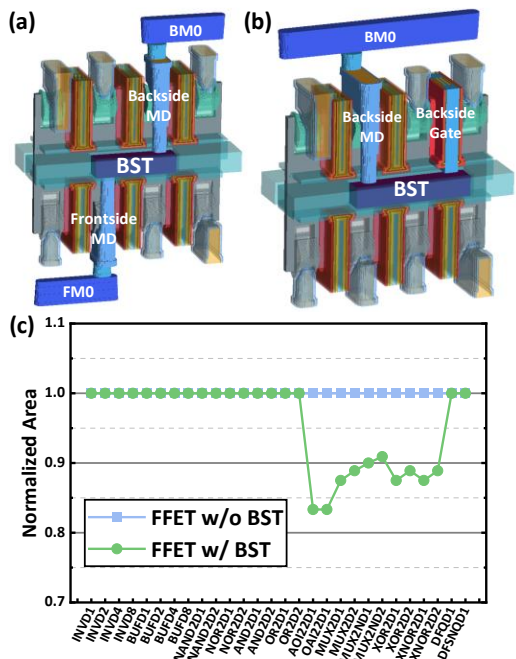

Fig. 5. (a) 3D structures of BST connecting frontside and backside S/D metals. (b) 3D structures of BST connecting backside S/D metal and backside gate. (c) Normalized area comparison between FFET with and without BST.

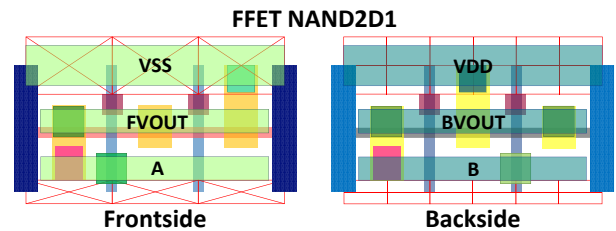

Fig. 6. Layout of FFET NAND2D1.

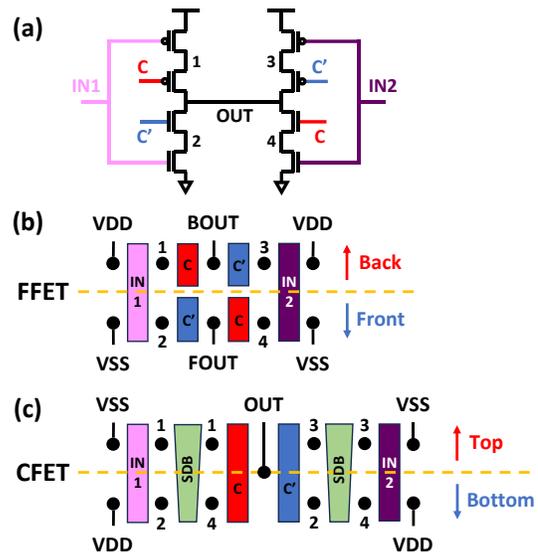

Fig. 7. (a) A typical subcircuit of $C^2MOS$. (b) FFET structure diagram with split gate design. (c) CFET structure diagram with common gate design. 40% area is wasted and more M0 signal tracks are needed.

## A. General Design Principles

For the FFET, we use a symmetric dual-sided std. cell design methodology. The N logic and P logic are formed separately on the wafer frontside and backside, then use the DM and GM to connect with each other. Unlike the CFET, this symmetric design can bring lower MOL parasitic RC and better routability. For cases with local routing difficulties, the FDM can be employed in addition to the routing with DM only.

On the other hand, due to the dual-sided routing rule [11][13][15], each FFET std. cell has the output pin placed on both the frontside and the backside. Differently, each input pin can be placed on either the frontside or the backside, depending on the cell requirement. Benefited from this design, the cell can drive the next stage's cells by connecting the output and input on the corresponding side flexibly.

Fig. 6 shows the layout of a typical 2.5T FFET std. cell: NAND2D1. Two nFETs are placed on the frontside while two pFETs on the backside. The VSS and VDD power rails are placed on opposite sides according to the type of transistors. The gates of nFET and pFET are connected by the GM as the input and two input pins are evenly distributed to both sides. The output is enabled by the DM and then picked up by the M0 on both sides. Note that the input pins can be freely adjusted for each side of wafer depending on the cell routing requirements.

## B. Split Gate Design

In the sequential logic circuit design, the clocked-CMOS ($C^2$MOS) is a must. The $C^2$MOS consists of a pair of CMOS and a pair of split gate transistors controlled by complementary signal (C and C'). Fig. 7(a) shows a typical application of $C^2$MOS. For the FFET, it can be easily designed thanks to the split gate design (Fig. 7(b)). However, in the monolithic CFET design, this subcircuit needs to be realized by common gate, featuring a huge 40% area penalty (Fig. 7(c)). What's worse, CFET needs the M0 to realize the complex interconnects while FFET does not, which brings more potential in the cell height shrinking for FFET.

Furthermore, in the combinational logic design, the $C^2$MOS and transmission gate (TG) are also widely used. Fig. 8(a) shows the layout of an XOR2D1 gate. The four transistors in the red box constitute a $C^2$MOS, where a split gate structure is used. The four transistors in the blue box constitute a TG. A common gate design is used for this TG and the GM is used to transmit signals between the two sides. Except the two transistors needed to constitute a TG, the other two unnecessary transistors are disabled by shorting their Sources and Drains.

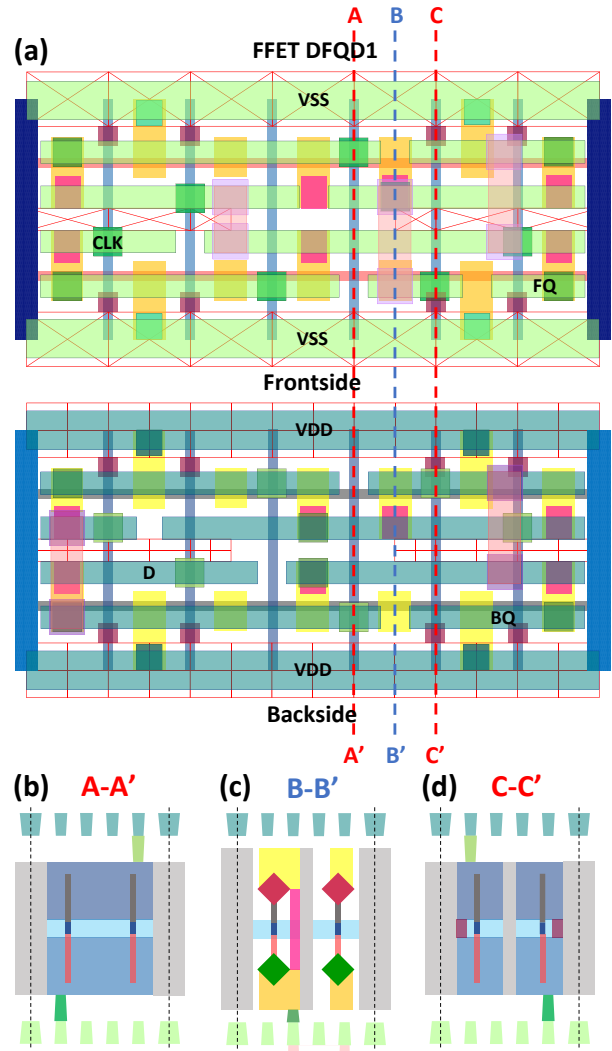

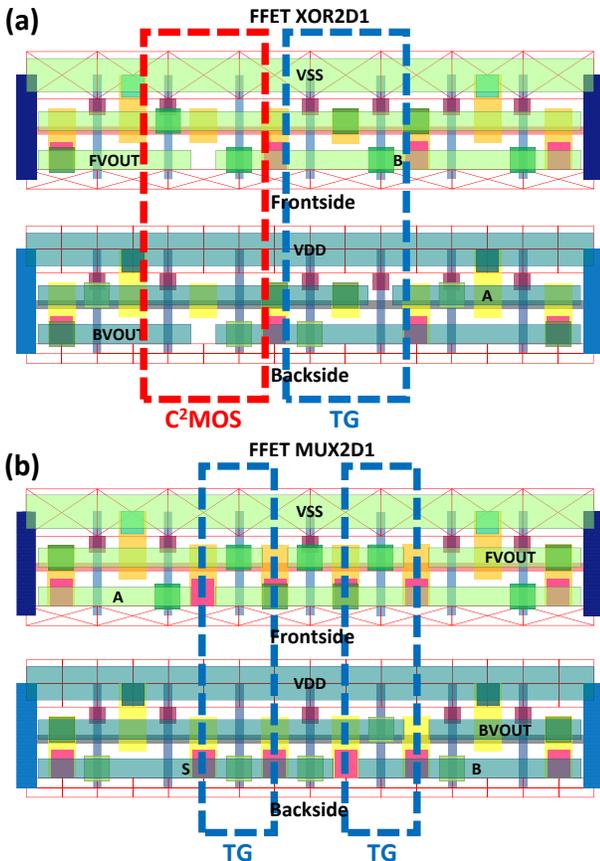

Fig. 8. (a) Layout of FFET XOR2D1. (b) Layout of FFET MUX2D1.

Fig. 9. (a) Layout of FFET DFQD1. (b) Cross-section of A-A'. The two gates on upper and lower rows are connected together while the gates on frontside and backside are not. (c) Cross-section of B-B'. Using M1 for cell interconnect. (d) Cross-section of C-C'. The two gates on frontside and backside are connected together while the gates on upper and lower rows are not.

To further illustrate the concept, fig. 8(b) shows the layout of a MUX2D1 gate. The transistors in the blue boxes constitute two TGs using the split gate design. The two transistors between them are used to transmit the S signal between the frontside and backside using GM, and they are both disabled by shorting the source and drain electrodes.

### C. Multi-row Design

For complex std. cells like the DFQ, more M0 signal tracks are needed for interconnects. As a solution, the multi-row design can be used to relax the local routing congestion [16]. First, the M1 can be used more often for cell interconnects to reduce the burden of M0 usage. What's more, the gates and S/D metals can be used for local interconnects between rows by removing the gate cut and S/D metal cut. Finally, more M0 signal tracks can be used in multi-row cells.

Fig. 9 shows a multi-row layout design of DFQD1 and its three cross-sections. Fig. 9(b) shows the cross-section of A-A'. The two gates on the upper and lower rows are connected by removing the gate cut in the middle so that the CLK signal can be transmitted between rows. Note that there's no GM between the frontside and backside gates. The frontside gates carry CLK signal while the backside gates carry CLK' signal, which forms split gate needed in the C$^2$MOS of DFQs. Fig. 9(c) shows the cross-section of B-B' and an M1 track is used to bridge two M0 signals. The C-C' cross-section is given in Fig. 9(d), showing two gates and GMs. In conclusion, with the multi-row and split gate design, we can freely choose the connection mode of the four gates. With multi-row and various S/D metal cuts, we can discretionarily decide the connection mode of the four S/D metals.

### D. An Example: AOI22 Design

The And-Or-Inverter21 (AOI21) and the And-Or-Inverter22 (AOI22) are common std. cells with complex cell interconnects and high pin density (Fig. 10(a-b)). The AOI21 can be easily designed according to the methods above. First, the drains of the nFET and pFET with input B (in following, they are shorted as B nFET and B pFET) are connected by a DM. Then a M0 line is used to join the drain of the A1 nFET (Fig. 10(c)).

However, the FFET AOI22 design is not that straightforward as shown in Fig. 10(d). The four drains of the A1 nFET, B1 nFET, B1 pFET and B2 pFET need to be connected as the output. However, by using the normal design, the drains between the frontside A1 & B1 and backside B1 & B2 are not in the same location on the layout, as they are vertically

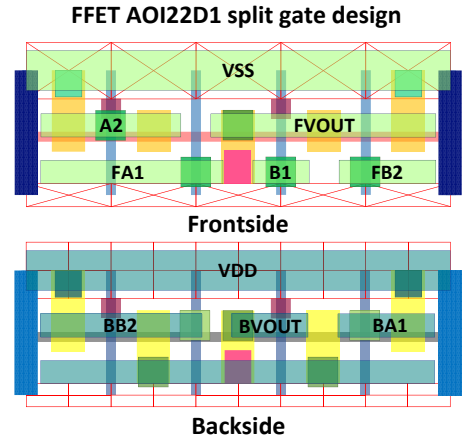

Fig. 11. Layout of FFET AOI22D1 with split gate design.

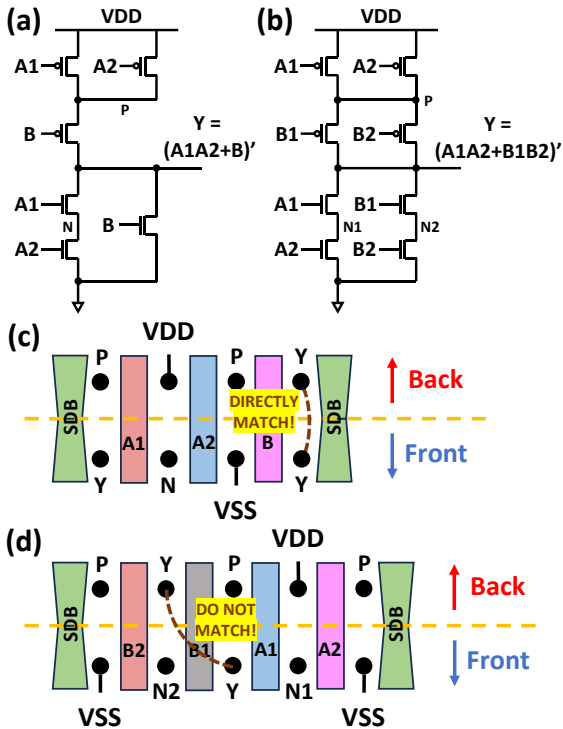

Fig. 10. (a) Schematic of AOI21. (b) Schematic of AOI22. (c) Structure diagram of AOI21 design. Frontside and backside output can be easily connected together by DM. (d) Structure diagram of AOI22 design. Frontside and backside output do not match.

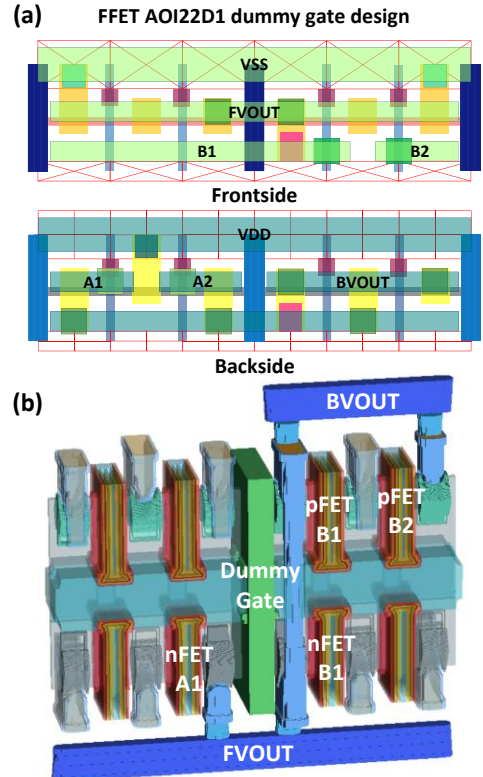

Fig. 12. (a) Layout of FFET AOI22D1 with dummy gate in the middle. (b) 3D structure of DG(DMmid) design. Four output nodes are connected together by DM and M0.

staggered. Thus, the DM cannot be used as in AOI21, failing the AOI22 design. Fortunately, there are several other options.
*1) Split Gate Design*

The first solution is utilizing the split gate design (Fig. 11). The position of the transistors is reassigned to make sure the drains of the A1 and B1 nFETs and B1 and B2 pFETs are vertically aligned. Then a DM is used to directly connect them together as the output. However, after relocating the transistors, some input signals would have to be fed into the cell from both frontside and backside independently, due to that there is no interconnect resource to combine those input signals.

Unfortunately, this could bring great challenges to the signal integrity. The two input signals on both sides may have a phase skew due to RC delays and the nFET and pFET with the same input signal may both be open, leading to a huge leakage path. As a result, this design is not preferred.
*2) Inserting Dummy Gates*

From above, the common-gated signal input is essential in the AOI22 design and new method of inserting dummy gates is introduced to enable it. As in Fig. 12, dummy gates (stripes with white dots) are inserted in the middle of the cell. Through this, the sources and drains of two adjacent transistors can be separated by the dummy gate. Next, all the four output nodes are placed next to the dummy gate and connected by the DM and M0. However, as the tradeoff, an extra CPP of area is wasted in this design.

There are two designs of AOI22 by inserting dummy gates. One design is shown in Fig. 12 where the dummy gate is the green block in the middle of the cell. This design is named DG(DMmid) because its DM is in the middle of the cell and next to the dummy gate. In another design, we switch the gates of B1 and B2. This way, the DM will come to the cell boundary. As a result, this design is called DG(DMside).
*3) FDM Design*

Apart from inserting dummy gates, the third approach is using the FDM design. The layout is shown in Fig. 13(a). Notice that there's no active fin within the far-left region where the FDM locates. Fig. 13(b) shows the 3D structure of the whole cell. By using the FDM, the frontside and backside's outputs can be connected and then picked up by the M0. With this design, only one pin is needed for each input signal, avoiding the issues in the multi-input design.

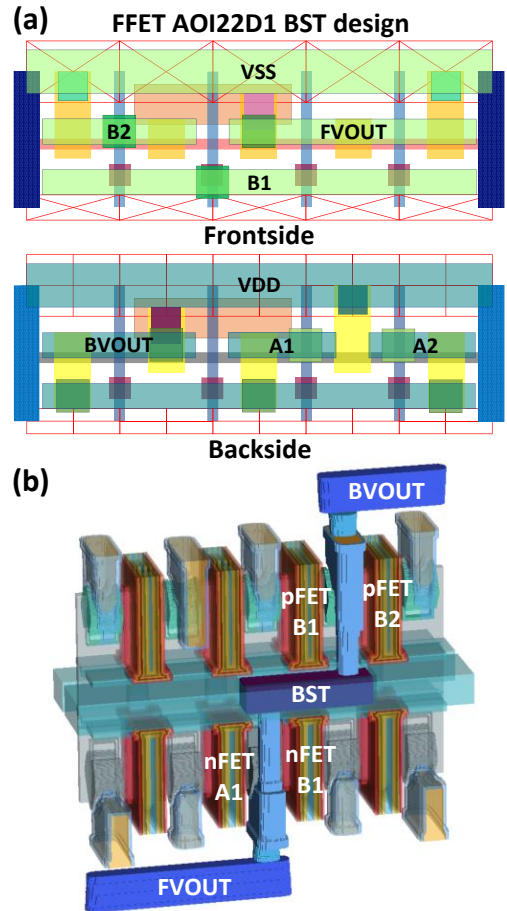

Fig. 14. (a) Layout of FFET AOI22D1 with BST. (b) 3D structure of BST design. Frontside and backside output are connected together by BST.

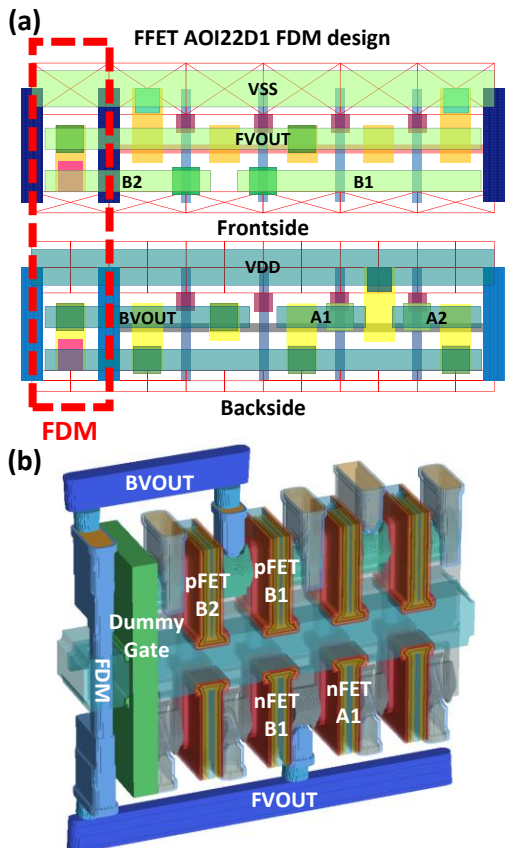

Fig. 13. (a) Layout of FFET AOI22D1 with FDM. (b) 3D structure of FDM_20 design. Frontside and backside output are connected together by FDM.

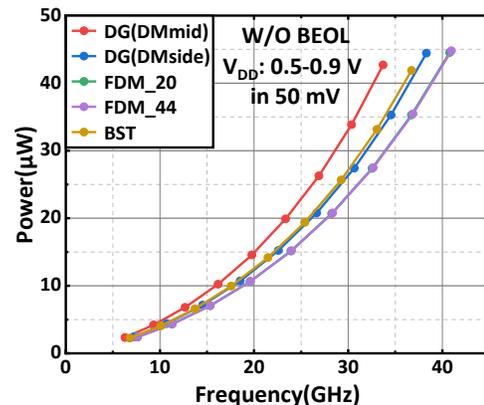

Fig. 15. Power-Performance curves of five different AOI22 designs.

Following the DM size in common cells, the width of DM used in the FDM design is also 20 nm. However, the width of DM can be increased to 44 nm, free of the constrains in normal DM due to the simple structure in this non-active region. By increasing the width of DM, the resistance drops but with capacitance penalty. To further reduce the output capacitance, the metal gate next to the FDM can be replaced by a dummy gate, reducing direct overlap between the gate and the FDM.

*4) BST Design*

The fourth design is using BST structure. The layout and 3D structure are shown in Fig. 14. In this design, the BST is used to connect the frontside and backside's S/D metals, for the output of the cell.

As indicated in the schematic, the BST greatly simplifies the AOI22 cell's signal routing without wasting area, but with more complicated processes.

*5) Comparison of Different Designs*

For the above, we have introduced 4 different designs for the AOI22 cell with 5 different layouts. Fig.15 shows the PP curves of 5 different designs. B1 is chosen as the input pin of the 15-stage FO3 RO and the $V_{DD}$ varies from 0.5 V ~ 0.9 V in 50 mV step. In terms of the FDM design, the long DM via is placed on the edge of the cell, with SDBs on both sides to reduce the overlap between the DM and active gates for less parasitic capacitance. As a result, the FDM design is better than the DG and the BST. To further optimize the FDM structure, the width of the DM is increased from 20 to 44 nm to reduce the resistance, delivering better PP results by using wider DM. As for the BST design, despite the high parasitic R&C caused by the buried metal, its smaller area reduces the M0 length and thus the intrinsic R&C. Overall, the BST design has the medium performance among all the designs.

For further comparison of delay and transition times of different designs, the DG(DMmid) is chosen as the reference and the other 4 designs' delay and transition times are compared in Fig.16, with 2.667 ps input net transition and 200 aF total output net capacitance at $V_{DD}$ = 0.7 V. The results are similar as the PP curves in Fig. 15, where FDM is the best, followed by BST and DG designs.

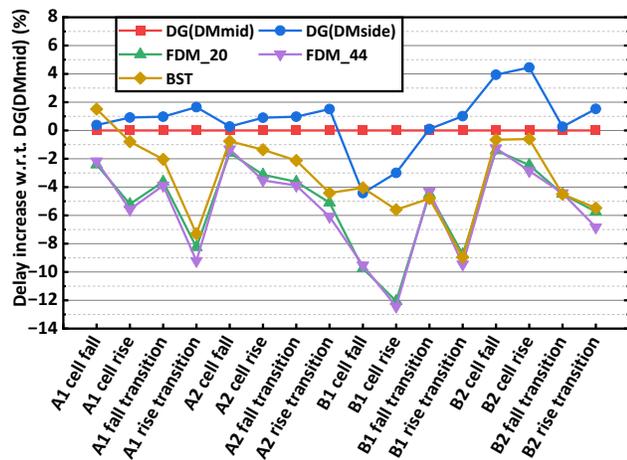

Fig. 16. Delay and transition time comparison between five different AOI22 designs.

## IV. CONCLUSION

In this work, we did a comprehensive study on 2.5T FFET's layout design. Compared with 3T CFET's design, 2.5T FFET has an average of 35.6% area reduction. Besides the multi-row design, split gate design and dummy gates insertion, other unique building blocks of FFET are examined, including the drain merge, gate merge, field drain merge and buried signal track. What's more, unique in FFFT, the symmetric design and dual-sided output pins are thoroughly evaluated. In addition, taking AOI22 as an example of complex cell design, multiple design options are discussed and the FDM design has an average of 1.93% and 5.13% lower delay and transition time than the BST and the dummy gates insertion methods, respectively. Finally, the R&C trade-off from varying the widths of the FDM is also investigated. For the first time, this work demonstrates multiple innovative approaches in designing FFET's layout, providing new insights for future's stacked transistor development.